\documentstyle[prd,aps,floats,psfig]{revtex}
\newcommand{\NP}[1]{Nucl.\ Phys.\ {#1}}

\newcommand{\PL}[1]{Phys.\ Lett.\ {#1}}

\newcommand{\PR}[1]{Phys.\ Rev.\ {#1}}
\newcommand{\PRL}[1]{Phys.\ Rev.\ Lett.\ {#1}}

\newcommand{\gsim}{\raise.3ex\hbox{$>$\kern-.75em\lower1ex\hbox{$\sim$}}}
\newcommand{\lsim}{\raise.3ex\hbox{$<$\kern-.75em\lower1ex\hbox{$\sim$}}}
\begin{document}
\draft
\input feynman
\renewcommand{\topfraction}{0.8}
\twocolumn[\hsize\textwidth\columnwidth\hsize\csname
@twocolumnfalse\endcsname
\preprint{hep-ph/9911297}

\title{\LARGE \bf
The  $\phi\rightarrow
\pi^+\pi^-$
decay within a chiral unitary approach}
\author {J. A. Oller$^1$, E. Oset$^2$ and J. R. Pel\'aez$^3$}
\address{$^1$ Forschungszentrum J\"ulich, Institut f\"ur
Kernphysik (Theorie). D-52425 J\"ulich, GERMANY.\\
$^2$ Departamento de F\'{\i}sica Te\'orica and I.F.I.C.
Centro Mixto Universidad de Valencia- C.S.I.C.\\
46100 Burjassot (Valencia), SPAIN.\\
$^3$ Departamento de F\'{\i}sica Te\'orica.
Universidad Complutense de Madrid.
28040 Madrid. SPAIN}
\date{March 8, 2000}
\maketitle
\begin{abstract}
  Starting from the Chiral Perturbation Theory
 Lagrangian, but keeping different
  masses for the  charged and neutral mesons ($m_u \neq m_d$), and using a
  previously developed
non-perturbative unitary scheme that generates the lightest
meson-meson resonances, we construct
$K \bar{K}\rightarrow K \bar{K}$
and $K \bar{K}\rightarrow \pi^+ \pi^-$ in the vector 
 channel. This allows us to obtain the kaon-loop
contribution to the $\phi-\rho$ mixing and study
the  $\phi\rightarrow\pi^+\pi^-$ decay.
The dominant contribution to this decay comes from
the $\phi\rightarrow\gamma\rightarrow\pi^+\pi^-$ process.
However, there can be large interferences with the subdominant contributions
coming from $\phi-\rho$ and $\phi-\omega$ mixing, or of these two
contributions among themselves. As a consequence, a reliable
measurement of $\phi\rightarrow\pi^+\pi^-$ decay could 
be used to differentiate
between some $\phi-\omega$ mixing scenarios proposed in the literature.
 \end{abstract}
\vskip2pc]
\section{Introduction}
    The $\phi$ decay into $\pi^+ \pi^-$ is an example of isospin
violation, since the $\phi$ has isospin $I=0$ and spin $J=1$, and it
would not couple to $\pi^+ \pi^-$ in the isospin limit,
which requires  $I+J=\hbox{even}$ (the decay into $\pi^0 \pi^0$ is
 forbidden in any case because the particles are identical). In addition,
it violates the OZI rule \cite{ozi} and hence it is subleading in
the large $N_c$ \cite{witten} expansion. The experimental
situation on this decay is rather confusing. There are two
old results:
$BR=(1.94+1.03-0.81)\times 10^{-4}$ from \cite{vaser}, and
$BR=(0.63+0.37-0.28)\times 10^{-4}$ \cite{golu},
with very different central values but whose errors
are so big to make them compatible.
Very recently,
two new, more precise,
but conflicting results have been reported
from the two experiments at the VEPP-2M in Novosibirsk:
the CMD-2 Collaboration reports
$BR=(2.20\pm0.25\pm0.20)\times 10^{-4}$ \cite{CMD2}
whereas the SND Collaboration \cite{SND} obtains
$BR=(0.71\pm0.11\pm0.09)\times 10^{-4}$.

On the theoretical side,
the common ground is based on the $\phi-\rho$
mixing  \cite{bramon,achasov,genz} to account for the strong part of the decay.
In addition, in ref. \cite{achasov} it has been pointed out
that the two-step
$\phi-\omega-\rho$ transition\footnote{As a matter of fact, 
this two-step process, just gives a 
contribution to the $\phi-\rho$ mixing. 
We will consider such resonant processs
 as the one that provides, by resonance saturation, 
the complementary local terms to 
the kaon loop contributions to $\phi-\rho$ mixing that we will
calculate later on.} 
 can give a relevant contribution
and that  other non-resonant processes, as a possible bare  
$\phi \rho \pi$ coupling, have to be considered in detail. It is remarkable,
in contrast with the OZI allowed
$\omega\rightarrow \pi^+\pi^-$
decay, that the electromagnetic $\phi\pi^+\pi^-$  coupling
via photon exchange $\phi - \gamma -\rho-\pi^+\pi^-$
provides the right order of magnitude \cite{bramon,genz}.

    Within Chiral Perturbation Theory (ChPT)
\cite{weinberg,xpt}, isospin breaking has recently gained interest, since
it is possible to take systematically
into account the corrections due
to the different $u$ and $d$ quark masses and to electromagnetic
effects. Examples of such
calculations are $\pi\pi$ scattering \cite{knecht}, some $\pi N$
 amplitudes  and the nucleon self-energy \cite{ulf1},
NN scattering \cite{ulf2} and the pionium atom \cite{gasser}.

Unfortunately, isospin violation in $\phi\rightarrow\pi^+\pi^-$ lies
 far away from
the ChPT applicability range, since it involves
the propagation of the pair of mesons around 1 GeV.
Nevertheless, new nonperturbative  schemes imposing unitarity and
still using the ChPT Lagrangian have emerged enlarging the
convergence of the chiral expansion \cite{ramonet,oop,ollerpaco}, for a review
see ref. \cite{review}.
Here we shall follow the work \cite{oop}, since it
provides the most comprehensive study of 
the different meson-meson scattering channels, including
resonances up to 1.2 GeV. In particular, this method yields a
resonance in the $I=0,J=1$ channel, which is related to the
$\phi$ and thus will allow us  to obtain an important
 contribution to $\phi\rightarrow\pi^+\pi^-$ due
to the charged and neutral meson mass difference.
We shall also consider electromagnetic
contributions at tree level as well as the contribution
due to the $\phi-\omega$ mixing.
These three contributions can have different kinds of cancellations
among themselves, depending on the $\phi-\omega$ mixing
scenario.

Some other theoretical uncertainties in our approach are unavoidable
since the results are rather sensitive to the
$L_i$ coefficients of the $O(p^4)$ ChPT Lagrangian and to the value of
$F_V$, which measures the coupling of a vector resonance with a photon.
We will not calculate the electromagnetic loop corrections since the
present ignorance of higher order counterterms makes their calculation
unfeasible. However, from refs. \cite{bramon,Lipkin,Renard} one expects the
meson--photon intermediate states to yield a contribution of, at most,
25$\%$ of that of kaon loops.

\section{Tree level contributions}

\subsection{The vector meson chiral Lagrangian}
In order to calculate the contribution of an intermediate photon
to $\phi\rightarrow\pi^+\pi^-$, we will use the vector meson chiral effective
Lagrangian presented in \cite{Ecker}, which
is written in terms of the SU(3) pseudoscalar meson matrix $\phi$
and the antisymmetric vector tensor field $V_{\mu\nu}$ defined as
  \begin{eqnarray}
\phi &=& \left(
  \begin{array}{ccc}
\frac{\pi^0}{\sqrt{2}}+\frac{\eta}{\sqrt{6}} & \pi^+ & K^+ \\
 \pi^-&-\frac{\pi^0}{\sqrt{2}}+\frac{\eta}{\sqrt{6}}& K^0\\
 K^-& \bar{K}^0&-\frac{2\eta}{\sqrt{6}}.
  \end{array}
\right), \nonumber \\
V_{\mu\nu} &=& \left(
  \begin{array}{ccc}
\frac{\rho^0}{\sqrt{2}}+\frac{\omega_8}{\sqrt{6}} & \rho^+ & K^{*\,+} \\
 \rho^-&-\frac{\rho^0}{\sqrt{2}}+\frac{\omega_8}{\sqrt{6}}& K^{*\,0}\\
 K^{*\,-}& \bar{K}^{*\,0}&-\frac{2\omega_8}{\sqrt{6}}
  \end{array}
\right)_{\mu\nu}.
\label{phiV}
 \end{eqnarray}
The latter is normalized such that
  \begin{equation}
\langle0\vert V_{\mu\nu}\vert P\rangle=\frac{i}
{M_R}[P_\mu\epsilon_\nu(R)-P_\nu\epsilon_\mu(R)],
\label{norm}
 \end{equation}
with $M_R$, $P$ and $\epsilon_\mu(R)$ the mass, momentum,  and
polarization vector of the vector field $R$.
Following \cite{Ecker},
 let us then consider the Lagrangian
 \begin{equation}
 {\cal L}_2[V(1^{-\,-})]=\frac{F_V}{2\sqrt{2}}
\langle V_{\mu\nu}f_+^{\mu\nu}\rangle+
\frac{i\,G_V}{\sqrt{2}}\langle V_{\mu\nu}u^\mu u^\nu\rangle,
\label{Lag}
 \end{equation}
where ``$\langle\;\rangle$" indicates the SU(3) trace and
  \begin{eqnarray}
u_\mu&=&i u^\dagger D_\mu U u^+ = u^\dagger_\mu \nonumber\\
 u^2 &=& U = \exp\left(\frac{i\sqrt{2}}{f}\phi\right)\nonumber\\
D_\mu U &=& \partial_\mu U- i\,e\,[Q,U]A_\mu,
\label{defs}
 \end{eqnarray}         
with $Q$ the quark charge matrix
  \begin{equation}
Q=\frac{1}{3}\,\hbox{diag}(2,-1,-1),
\label{Q}
 \end{equation}
and $A_\mu$ the electromagnetic field.
As usual, $f$ is the pion decay constant in the chiral limit (we take
$f\simeq f_\pi=92.4$ MeV) and the $f^{\mu \nu}_+$ and $F^{\mu \nu}$
 tensors are defined as
\begin{eqnarray}
f^{\mu \nu}_+ &=& u F^{\mu \nu} u^\dagger + u^\dagger F^{\mu \nu} u,\nonumber\\
F^{\mu \nu}&=&e\,Q\,(\partial^\mu A^\nu-\partial^\nu A^\mu),
\label{fplus}
\end{eqnarray}

In order to introduce the $\phi$ and $\omega$ states,
we extend SU(3) to U(3) and substitute
\begin{equation}
\nonumber
V_{\mu\nu}\rightarrow  V_{\mu\nu} + (\omega_{1})_{\mu\nu}
\frac{I_3}{ \sqrt{3}} ,
 \end{equation}
where $I_3$ is the diagonal $3\times3$ matrix and $\omega_1$ is
the lightest singlet vector resonance. 
Hence, by imposing ideal mixing between $\omega_1$ and $\omega_8$
  \begin{eqnarray}
  \frac{2}{\sqrt{6}}\omega_1+\frac{1}{\sqrt{3}}\omega_8= \omega^{ideal},
\nonumber \\
\frac{1}{ \sqrt{3}}\omega_1-\frac{2}{\sqrt{6}}\omega_8= \phi^{ideal},
\label{mixing}
 \end{eqnarray}
Unless otherwise stated, in the following
we will refer to these states simply as $\omega$ and
$\phi$, although
it should be kept in mind that we are referring 
to their respective ideal states. Finally,
the $\phi$ and $\omega$ can be introduced
into the chiral notation by replacing
in  eq.(\ref{Lag}) the
 $V_{\mu\nu}$ tensor by
  \begin{equation}
\tilde{V}_{\mu\nu} = \left(
  \begin{array}{ccc}
\frac{\rho^0}{\sqrt{2}}+\frac{\omega}{\sqrt{2}} & \rho^+ & K^{*\,+} \\
 \rho^-&-\frac{\rho^0}{\sqrt{2}}+\frac{\omega}{\sqrt{2}}& K^{*\,0}\\
 K^{*\,-}& \bar{K}^{*\,0}&\phi
  \end{array}
\right)_{\mu\nu}.
\label{Vtilde}
 \end{equation} 

  The convention of signs of eq.(\ref{defs})
agrees with a more standard one if
 we take $e$ negative in all the Lagrangians, as we shall do in what
 follows. The vertex function $\phi\rightarrow\gamma$,
resulting from eq.(\ref{Lag}), is
\begin{equation}
i\,{\cal L}_{\phi\gamma}\rightarrow -i\,\frac{\sqrt{2}}{3}\,\vert
e\vert\, F_V \,M_\phi \,\epsilon^\mu(\phi)\, \epsilon_\mu(\gamma),
\label{Lphigamma}
\end{equation} 
 and to the same order than eq.(\ref{Lphigamma}) the Lagrangian
giving the coupling of the photon to the
 pions is
\begin{equation}
i\,{\cal L}_{\gamma\pi^+\pi^-}=\vert
e\vert\left(\pi^-\partial^\mu\pi^+-\pi^+\partial^\mu\pi^-\right)A_\mu
\label{Lgammapipi}
 \end{equation}

    With these ingredients we can write the contribution of the
 Feynman diagram of fig.1,  which is given by
\begin{equation}
i e^2 \frac{\sqrt{2}\,F_V}{3\,M_\phi}
 \epsilon^\mu(\phi) (p_{\pi^+}-p_{\pi^-})_\mu\,F(M_\phi^2),
\label{Lphipipi}
 \end{equation}
where $F(q^2)$ is the pion electromagnetic form factor,
which at the $\phi$ mass is given by $F(M_\phi^2)=-1.56+i\,0.66$
  \cite{ffexp}.
  
This can be compared with the coupling of the $\phi$ to  $K^+\,K^-$,
or $K^0\,\bar{K^0}$, which can be obtained from the
$G_V$ term in eq.(\ref{Lag}) and reads
\begin{eqnarray}
i\,{\cal L}_{\phi K^+ K^-}&\rightarrow&-i\, g_{\phi K^+ K^-} \,
 \epsilon^\mu(\phi) \,(p_{K^+}-p_{K^-})_\mu\,,
\label{Lphikk}
\\\nonumber
g_{\phi K^+ K^-}&=&\frac{s\,G_V}{\sqrt{2}\,f^2\,M_\phi}\,.
\end{eqnarray}
 The $\phi\rightarrow K^+ K^-$ width
is then given by
\begin{equation}
\label{Gphikk}
\Gamma_{\phi K^+ K^-}=\frac{p_{K^+}^3}{6\pi\,M_\phi^2}\,
g_{\phi K^+ K^-}^2,
\end{equation}
which, using its experimental value \cite{PDG},
 provides $G_V=54.3$ MeV
(to compare with $G_V=53\,\hbox{MeV}$, from the study of
the pion EM radius \cite{xpt,Ecker}).

  By analogy to eq.(\ref{Lphikk}), eq.(\ref{Lphipipi})
provides a $\phi\pi^+\pi^-$ coupling
\begin{equation}
g_{\phi \pi^+ \pi^-}^{(\gamma)}
=-\frac{\sqrt{2}}{3}\, e^2 \frac{F_V}{M_\phi}
\,F(M_\phi^2),
\label{g8phikk}
\end{equation}
and eq.(\ref{Gphikk}), substituting $g_{\phi K^+K^-}$
by $g_{\phi \pi^+ \pi^-}$
 and $p_{K^+}$ by $p_{\pi^+}$,
provides the tree level electromagnetic contribution to the
$\phi\rightarrow \pi^+ \pi^-$ decay width.
With a value of $F_V= 154$ MeV from the $\rho\rightarrow e^+e^-$ decay
\cite{Ecker} this contribution alone would yield
$BR(\phi\rightarrow \pi^+ \pi^-)=1.7 \times 10^{-4}$,
a value 
compatible with the experiment of ref. \cite{CMD2}, within errors.

\subsection{Comparison with $\omega\rightarrow \pi^+\pi^-$}

It may seem surprising that $g_{\phi \pi^+ \pi^-}^{(\gamma)}$ already
provides the correct order of magnitude of the
$\phi\rightarrow \pi^+ \pi^-$ decay, since, in contrast,
it is well known \cite{gatto,shifman} that the tree level photon contribution
 $\omega \rightarrow \gamma \rightarrow \rho \rightarrow \pi^+\pi^-$,
represents a negligible amount of the $\Gamma(\omega \rightarrow \pi^+\pi^-)$.

However,  the case of the $\omega$
is radically different from ours and  can be well understood from
$\rho-\omega$ mixing. We will now calculate this effect
 making use of an effective chiral Lagrangian
and large $N_c$ arguments \cite{urech}. Indeed, from this reference, the
$\rho-\omega$ mixing can be represented as
\begin{equation}
\label{urech}
i\,{\cal L}_{\rho\omega}\rightarrow\,i\,
\tilde\Theta_{\rho \omega} \epsilon_\rho \cdot
\epsilon_\omega,
\end{equation}
with
\begin{equation}
\label{tecta}
\tilde\Theta_{\rho \omega}=\frac{s}{M_V^2}\left[-(m_{K^0}^2-m_{K^+}^2)+
(m_{\pi^0}^2-m_{\pi^+}^2)+\frac{1}{3}
F_V^2 e^2\right].
\end{equation}
 Note that $M_V$ is the mass of
the vector octet in the chiral limit $M_V\approx M_\rho$ \cite{Ecker}. We have
also made use in eq. (\ref{urech}) of eq. (\ref{norm}), thus turning to
the usual vector notation. At lowest order in ChPT  \cite{xpt}
the first two terms in eq. (\ref{tecta}) arise from the quark mass difference,
and the third one is of electromagnetic
origin from the exchange of a photon
between the $\rho$ and the $\omega$. It is straightforward to see that the
electromagnetic contribution only amounts to a 14$\%$ of that due
to quark mass differences.

Contrary to the $\phi$ case, the $\rho-\omega$ mixing is OZI allowed and
leading in
large $N_c$, as can be seen from eqs. (\ref{urech}) and (\ref{tecta}). In fact,
this term
is of the same order than
the free Lagrangian, both in the $1/N_c$ and in chiral countings (this is more
clearly seen in the tensor notation).

In addition, there is a kaon loop contribution, fig.2, which,
from ChPT, is expected to be of the same order of magnitude
 than the electromagnetic
contribution. Evaluating the diagram of fig.2 one has:
\begin{equation}
\label{loopkaon}
\tilde\Theta_{\rho\omega}^{\footnotesize Kaon-loops} =
\frac{G_V^2}{f_\pi^4}\frac{s^2}{(4\pi)^2 M_V^2}
\left[L(s,m_{K^+})-L(s,m_{K^0}) \right].
\end{equation}
where, once again, we have used eq. (\ref{norm}) to present our results in the
vector notation. The  $L(s,m)$ loop function, in
the usual ChPT $\overline{MS}-1$ scheme, is:

\begin{eqnarray}
\label{loopfunction}
L(s,m)&=&\frac{m^2-s/6}{3}-\frac{m^2}{6}\log \frac{m^2}{\mu^2}-\nonumber\\
&& \frac{s-4 m^2}{12} \left[ 1
-\log \frac{m^2}{\mu^2}-\sigma \log\frac{\sigma+1}{\sigma-1}\right] \nonumber \\
\sigma&=& \sqrt{1-\frac{4 m^2}{s}},
\end{eqnarray}
where $\mu$ is the dimensional regularization scale. In order to estimate
the eq. (\ref{loopkaon})
contribution at $s=M_\rho^2$, we use
the natural value $\mu=\Lambda_{\hbox{ChPT}}  \approx M_\rho$.
The results depend on the regularization scale but
they provide a good estimate of the order of magnitude, 
as we shall see later on, when we will reevaluate this contribution 
within the chiral unitary approach.

At this point we are ready to compare all contributions:

\vskip 0.3cm

{\setlength{\tabcolsep}{.1mm}
\begin{tabular}{lcr}
Quark mass differences from eq.(\ref{tecta})& = & $-5221.6$ MeV$^2$ 
\nonumber \\
EM contribution from eq.(\ref{tecta})& = & 725.1 MeV$^2$ \nonumber \\
Kaon loops from eq.(\ref{loopkaon})& = &$-130$ MeV$^2$
\end{tabular}}

\vskip 0.3cm

Hence, the $\rho-\omega$ mixing is dominated by the OZI allowed strong contribution
due to quark mass differences, which is leading
both in the large $N_c$ and chiral countings.
In addition, the kaon loops are smaller than the
electromagnetic contribution although with a large destructive
interference between them (for $G_V=65$ MeV, which is the value
needed to reproduce $\Gamma(\rho \rightarrow \pi^+\pi^-)$ from eq. (\ref{Lag}),
the estimate of the kaon loop contribution would be $-$190
MeV$^2$). We will find again this large destructive interference between the
kaon loops and the electromagnetic contribution when considering the $\phi$ resonance.

In summary, the fact that the purely
electromagnetic contribution already provides a reasonable 
order of magnitude for
the $\phi\rightarrow\pi^+\pi^-$ decay, is due to the absence of the OZI
allowed contribution, which makes the $\omega\rightarrow\pi^+\pi^-$
decay comparatively much larger. Note that such contribution 
is missing in the OZI
violating, large $N_c$ subleading, $\phi-\rho$ mixing.
The fact that $\omega\rightarrow\pi^+\pi^-$ is much larger 
than the $\phi\rightarrow\pi^+\pi^-$ dominant contribution is very
relevant since, through the $\phi-\omega$ mixing, it provides 
an additional mechanism that has to be taken into account
in the complete calculation of $\phi\rightarrow\pi^+\pi^-$,
and that we analyze next.

\subsection{The ``two step'' $\phi-\omega-\rho$ mechanism}

As a matter of fact, the physical $\phi$ and $\omega$ states are not
the ideal ones defined in eq.(\ref{mixing}), but instead
\begin{eqnarray*}
\omega \simeq \omega^{ideal}
-\delta_V \phi^{ideal}
\\ 
\phi\simeq \delta_V\omega^{ideal}+\phi^{ideal}
\end{eqnarray*}
In the literature there is a general agreement on 
$\vert \delta_V\vert\simeq 0.05$, but, apart from conventions, 
not on its sign \cite{achasov}. Its
contribution to the $\phi\pi^+\pi^-$ effective coupling
is obtained from fig.3 as follows:
\begin{eqnarray}
g^{(\phi \omega)}_{\phi\pi^+\pi^-}&=&-
 \frac{M_\phi^2
G_\rho}{M_\rho f^2}\frac{
\tilde\Theta_{\rho\omega}(M_\phi) }{M_\phi^2-M_\rho^2
+i M_\rho \Gamma_\rho}\nonumber \\
&&\frac{\tilde\Theta_{\phi \omega}(M_\phi) }
{M_\phi^2-M_\omega^2+iM_\omega
  \Gamma_\omega}.
\end{eqnarray}
We have already obtained $\tilde\Theta_{\rho\omega}$, although here it
has to be evaluated at $\sqrt{s}=M_\phi$. Still, the dominant
contribution comes from the quark mass differences. The kaon loop
contribution cannot be calculated using eq.(\ref{loopkaon}), since 
that formula is not unitary. We will see later, how this number can be 
obtained from the chiral unitary approach, and again it is of the
order of $200\,\hbox{MeV}^2$, and therefore numerically irrelevant for the
following discussion. 

The new $\tilde\Theta_{\phi \omega}$ parameter can be obtained from
  the literature. Nevertheless, its imaginary part can 
be obtained from unitarity. The most relevant intermediate states
are $K\bar{K}$ and three pions. In the first case
the couplings to $\phi$ and $\omega$
are completely determined by the vector resonance Lagrangian. 
However, the imaginary part contribution of three pion
intermediate states has some model dependence \cite{achasov},
mostly through the $g_{\phi\rho\pi}$ coupling.

 We consider now two different scenarios for the $\phi-\omega$ 
mixing which illustrates to some extent
the uncertainties that are found in the literature with respect to this issue:
\begin{itemize}
\item ``Weak mixing'' scenario \cite{achasov}, where
$\hbox{Re}\,\tilde\Theta_{\phi\omega}=0$ and
$g_{\phi\rho\pi}=0.78 \,\hbox{GeV}^{-1}$.
\item ``Strong mixing'' scenario  \cite{achasov}, where
$\hbox{Re}\,\tilde\Theta_{\phi\omega}=20000$ 
to $29000\, \hbox{MeV}^2$ and
$g_{\phi\rho\pi}=0$.
\end{itemize}
which will therefore appear as different cases in our final result.

Up to now we have just concentrated on the tree level diagrams
of the $\phi\pi^+\pi^-$ decay. There are, however, important contributions
from kaon loops that we will analyze in the next sections, whose 
calculation is the main novelty of this work.

\section{Direct kaon loop contribution to $\phi-\rho$ mixing}

\subsection{Introduction}
The pure strong interaction chiral Lagrangian gives a contribution to the
$\phi\rightarrow \pi^+ \pi^-$ decay if the charged and neutral meson
 masses are different, otherwise it
would be forbidden.

 For instance, from eqs.(\ref{Lag}) and
(\ref{Vtilde}) there is no direct $\phi\pi^+ \pi^-$ coupling.
However, we can generate a non vanishing $\phi\rightarrow \pi^+ \pi^-$
 transition when keeping different masses for the
charged and neutral kaons in the loops
 of fig.4, which do not violate the OZI rule, 
although they are subleading in large $N_c$. In
 fact, these diagrams are expected to give the
main strong interaction contribution to $\phi \rightarrow \pi^+\pi^-$ due to
intermediate states. For instance, the $\phi$
couples much more strongly to $K\bar{K}$ than to $3\pi$,
as it is clear from the fact that
$\Gamma(\phi\rightarrow 3\pi)/\Gamma(K\bar{K})\approx 1/5$, although three pions
are kinematically much more favored than two kaons
\footnote{We will address in subsection III.D
the problem raised in refs. \cite{isgur,isgur2},
relative to the contributions of more
massive virtual intermediate states}.
 
Note that in the evaluation of the diagrams of fig.4 the 
$K\bar{K}\rightarrow \pi^+\pi^-$ amplitude can receive important contributions 
from the $\omega$ or $\rho$ exchange. In the first case, the $\omega$ 
couples to the $\rho$ once again, and therefore is included
in the $\phi-\omega-\rho$ mixing contributions. 
Thus, in the following we will concentrate on the 
evaluation of these kaon-loop contributions to 
the direct $\phi-\rho$ mixing, that is, 
we will consider only the exchange of the $\rho$ in the $K\bar{K}\rightarrow 
\pi^+\pi^-$ $I=0$ P-wave amplitudes appearing on fig.4.

An estimation of the imaginary part of this contribution to the
diagrams 
in fig.4 is 
straightforward using the vector meson chiral Lagrangian. The sum 
of the diagrams does not vanish due to the different masses of the 
charged and neutral 
kaons. The $\phi \rightarrow \pi^+\pi^-$ branching ratio that would be 
obtained  taking into account  just
this contribution is already of the order of magnitude
of the experimental results, given the large 
uncertainty on the data.

Yet, this estimate does not take into account corrections to
$K \bar{K}\rightarrow \pi^+ \pi^-$
due to isospin violation. In addition, the real part of the loop
remains ambiguous since it requires the knowledge of higher order
contributions than those given by eq.(\ref{Lag}), that is, counterterms
to absorb loop divergences. Furthermore, even when we have such
counterterms, the chiral expansion is only expected to work at energies
which are below the $\phi$ mass.

\subsection{Resonances and the IAM}

We present here a method which deals simultaneously with all these
problems in order to 
extract the aforementioned kaon loop contributions. The method  exploits the
information of ChPT up ${\mathcal{O}}(p^4)$, by relying on the expansion of the
$T^{-1}$-matrix. The technique starts from the $O(p^2)$ and
$O(p^4)$ ChPT Lagrangian and uses the
inverse amplitude method (IAM) in coupled channels.
Unitarity provides for free the imaginary part of $T^{-1}$, and then
a chiral expansion is done for Re$\,T^{-1}$, which, in the present case,
has a larger radius of convergence than $T$ itself.
This approach has been applied in the
isospin limit with remarkable results:
with just one channel \cite{ramonet} it nicely describes the $\sigma$,
$\rho$ and $K^*$ regions, amongst
others, in $\pi^+\pi^-$ and $\pi K$ scattering.
When generalized to coupled channels \cite{oop,ollerpaco} it
also describes meson-meson scattering with all the associated
resonances up to about 1.2 GeV. A more general approach is
used in \cite{oo} by means of the N/D method, in order to include
the exchange of some preexisting resonances explicitly,
which are then responsible for the values of the
fourth order chiral parameters.

The $T$ amplitude is defined in terms of the partial waves as
\begin{equation} 
T = \sum_J (2J+1)\,T_J(s)\,P_J(\cos\theta),
\nonumber
\end{equation}
In what follows we will refer to $T_J$ simply as $T$. Within the
coupled channel formalism, the IAM partial wave amplitude
is given by the matrix equation
\begin{equation} 
T=T_2\,[T_2-T_4]^{-1}\,T_2,
\label{IAM}
\end{equation} 
where $T_2$ and $T_4$ are  $O(p^2)$ and $O(p^4)$ ChPT
partial waves, respectively. In principle, $T_4$
would require a full one-loop calculation, but it was shown in
\cite{oop} that, at the phenomenological level, it can be well approximated by
\begin{equation} 
\hbox{Re}\,T_4\simeq T_4^P+T_2\,\hbox{Re}G\,T_2
\label{Ret4}
\end{equation}  
where $T_4^P$ is the tree level polynomial contribution coming
from the ${\cal L}_4$
chiral Lagrangian and $G$ is a diagonal matrix $diag(g_1,g_2,g_3)$,
where $g_i$ is the loop function
of the intermediate two meson propagators, which we give in the appendix. 
In  \cite{oop}
the loop integrals are regularized by means of a momentum cut-off, $q_{max}$, in the loop
three-momentum. The relation between this cut-off and the
dimensional regularization scale $\mu$,
normally used in ChPT, is also  given in that paper.

We have also taken advantage to  correct a small error detected in \cite{oop} in
the $K^+K^- \rightarrow K^0\bar{K}^0$ amplitude, whose complete expression
in the isospin limit is given in the appendix.
We have also reconducted a fit to 
the data including those on ($\delta_{00} - \delta_{11})$, which are
well determined from \cite{roselet}. The fit of the
phase shifts and inelasticities is carried out here
in the isospin limit, as done in \cite{oop}. There are several sets
of $L_i$ coefficients which  give rise to equally acceptable fits.

As it can be seen in fig.5, there are several plots for which there
are incompatible sets of data. This is particularly evident for
the $\delta_{00}$ data both in $\pi\pi\rightarrow\pi\pi$ and $K \bar K$,
in the inelasticity $\eta_{00}$, and in the $\delta_{0\,1/2}$ phase shifts.
As a consequence, although we have performed a $\chi^2$ fit of the data using
MINUIT \cite{MINUIT},
the resulting $\chi^2$ per degree of freedom is not really very
meaningful,
since the $L_i$ values depend on the estimate of
the systematic error of each experiment, which is not given in many original
references.
In addition, due to the fact that we have eight parameters,
there are several $\chi^2$ minima, which yield very similar values of $\chi^2$
for rather different values of some chiral parameters. Which one is the real
minimum depends on how we add the systematics. For that reason we have preferred
to give several sets of coefficients, which yield $\chi^2/d.o.f<2$
when
assuming a 3\% systematic error  added in quadrature to the statistical
error quoted by each experiment.
 
We write in table 1 the
different sets of chiral parameters and we show their corresponding results for
the phase shifts and inelasticities in fig.5.
We can see that the small differences
in the results appear basically only in the
$a_0(980)$ and $\kappa(900)$ resonance regions,
where data have also larger errors or are very scarce.

 Although the tadpoles and loop terms
in the crossed channels were neglected and reabsorbed into
$L_i$ redefinitions  \cite{oop} when we use eq.(\ref{Ret4}),
these coefficients are still
close to those of standard ChPT (see Table I). Consequently, it seems that
this simplifying approximation has
a small effect in the relevant energy region, not spoiling also the standard
low energy ChPT results.

One of the side consequences of the approach was the generation
of a resonance around 1 GeV in the $I=0$ and $J=1$ channel, which
only couples to $K \bar K$. Actually, it has a zero width,
since its mass is below the $K \bar{K}$ threshold. One is tempted to
associate this state to the $\phi$ meson, however, we can only
relate it with
the octet part $\omega_8$, which, by mixing with a singlet
generates the $\phi$ and the $\omega$. This can be
easily understood since the singlet in this
channel, $\omega_1$, which is symmetric in the SU(3) representation,
does not couple to two mesons because their spatial wave function is
antisymmetric.  Since only
two meson states were considered in  \cite{oop,ollerpaco},
$\omega_1$ does not appear in the  IAM, and
the resonant state found in that channel
can only be related to $\omega_8$. However,
 we will see next that we can still
exploit the properties of the $w_8$ pole in
order to study the decays of the $\phi$ resonance.

\subsection{Extracting the $\phi\pi^+\pi^-$ coupling from the IAM}
 
Let us then  turn to the case of interest for this work:
the evaluation of the $J=1$ $K\bar{K}\rightarrow K\bar{K}$ and 
 $K\bar{K}\rightarrow \pi^+\pi^-$ amplitudes around the mass of the
$\omega_8$. Now we are breaking
isospin explicitly by keeping different the charged and neutral meson masses,
while keeping the $L_i$ obtained from the previous fits
to meson-meson scattering in the isospin limit.
In addition, we are dealing with three two-meson states:
$K^+ K^-$, $K^0 \bar{K}^0$ and $\pi^+ \pi^-$, that we will call
1, 2 and 3, respectively. The amplitude is a $3\times 3$ matrix
whose elements we will denote as $T_{ij}$
(for instance, $T_{13}$ stands for
the $J=1$ $K^+ K^-\rightarrow \pi^+ \pi^-$ amplitude).
The $T_2$ and $T_4^P$ amplitudes used in the present work and calculated
in the isospin breaking case, are collected in the appendix.

Once the amplitudes are unitarized with the IAM,
one observes the presence of two poles, one corresponding to the
$\rho(770)$ and the other one to the $w_8$ resonance. It is interesting to note
that the
$w_8$ pole appears with a mass around 910 MeV, very close to the value 930 MeV
predicted by the quadratic or linear $SU(3)$ mass formulae \cite{okubo} for the $w_8$ mass.
In the following, we will denote by $\Omega_8$ the resonance pole that we
have obtained in our approach corresponding to the $w_8$ resonance. The motivation for
this change of notation is the lack of the $3\pi$ state in our model since this contribution
can be particularly relevant in order to study certain properties of the $w_8$ resonance.
For instance, the $3 \pi$ couplings of the $w_8$ and $w_1$ according to eq. (\ref{mixing}) add
for the $\omega$ giving rise to the $w\rightarrow 3\pi$ coupling and almost cancel each other
in the case of the $\phi$, $|g_{\phi \rightarrow 3\pi}| << |g_{\omega \rightarrow 3\pi}|$.

In order to evaluate the kaon loop contribution to the 
$\phi \pi^+\pi^-$ coupling via direct $\phi-\rho$ mixing, 
we first study the
$\Omega_8\pi^+\pi^-$ coupling. We thus evaluate the
$K^+K^- \rightarrow K^+ K^-$ amplitude ($T_{11}$) and the
$K^+ K^- \rightarrow\pi^+ \pi^-$
amplitude ($T_{13}$) near the pole of the $\Omega_8$ resonance. Close to the
$\Omega_8$ pole the amplitudes obtained numerically are then dominated by the
exchange of this resonance, represented diagrammatically in fig 6.

By considering couplings like those in eq.(\ref{Lphikk}) for the $\Omega_8$
to $K^+ K^-$ and $\pi^+ \pi^-$, these two amplitudes, once projected in 
the $J=1$ channel, and close to the $\Omega_8$
pole, are given by
\begin{eqnarray}
T_{11}&=&g_{\Omega_8 K^+ K^-}^2\frac{1}{s-M^2_{\Omega_8}}\,\frac{4\,
p_K\, p_{K'}}{3}\nonumber\\
T_{13}&=&g_{\Omega_8 K^+ K^-}\,g_{\Omega_8 \pi^+ \pi^-}
\frac{1}{s-M^2_{\Omega_8}}\,\frac{4\,
p_K\, p_{\pi}}{3}.
\label{t11t13}
\end{eqnarray}
where $p_i$ is the modulus of the center of mass 
three-momentum of the $i$ particle.
The diagram of fig.6b can be interpreted as providing an effective
 strong $g^{(s)}_{\Omega_8 \pi^+ \pi^-}$ coupling.

By looking at the residues of the $T_{11}$ and
$T_{13}$ amplitudes in the $\Omega_8$ pole we can get
$g_{\Omega_8 K^+K^-}\, g_{\Omega_8 K^+K^-}$ and
$g_{\Omega_8 K^+ K^-}\, g_{\Omega_8 \pi^+\pi^-}$. Thus, defining
   \begin{equation}
 \label{Qij}
 Q_{ij}=\lim_{s\rightarrow
M^2_{\Omega_8}} (s-M^2_{\Omega_8})\frac{3\,T_{ij}}{4\,p_i\,p_j}
 \end{equation}
   we obtain
   \begin{equation}
\label{fracgs}
\frac{g_{\Omega_8 \pi^+ \pi^-}}{g_{\Omega_8 K^+ K^-}}=
\frac{Q_{13}}{Q_{11}}.
 \end{equation}

   In eq.(\ref{Qij}) the $T_{11},T_{13}$ amplitudes
 have a large $\rho$ exchange
 background, which can be eliminated using
the residue of the $\Omega_8$ pole obtained via
eq.(\ref{Qij}). Yet, numerically this background can be eliminated to a large
extend by using the isospin zero combination
 $-(K^+K^-+K^0\bar{K}^0)/\sqrt{2}$ in the initial state. Hence, the $g_{\Omega_8 \pi^+\pi^-}$ is
 more efficiently evaluated by means of the combination
\begin{equation}
\label{fing}
\frac{g_{\Omega_8 \pi^+ \pi^-}}{g_{\Omega_8 K^+ K^-}}=
\frac{Q_{13}+Q_{23}}{Q_{11}+Q_{21}}.
\end{equation}
We have checked numerically that the 
$g_{\Omega_8 K^+K^-}$ and the $g_{\Omega_8 K^0\bar{K}^0}$ couplings 
have the same 
value in our approach.
Since we are interested in $\phi$, we still have to make the connection between
the $\Omega_8\pi^+\pi^-$ and $\phi\pi^+\pi^-$ couplings.
Indeed, we have explicitly checked that, when removing the rescattering
resummation implicit in
the IAM (by setting $G=0$, see  eq. (\ref{Ret4})), the ratio in eq. (\ref{fing}) becomes between
one and two orders of magnitude smaller.
Even more, this drastic reduction in the $\Omega_8 \pi^+\pi^-$
coupling is also obtained when
making $G=diag(0,0,g_3)$, that is, when only removing the kaon loops.
Therefore the
$\Omega_8$ decays to $\pi^+\pi^-$ mainly through the mechanism shown in fig.4
(replacing the $\phi$ by the $\Omega_8$).
This observation allows us to find the kaon loop contribution to
$\phi\rightarrow\pi^+\pi^-$ that we are looking for,
through the same mechanisms of the $\Omega_8$,
since the only difference will be the initial $\Omega_8K\bar K$ and $\phi K\bar K$
couplings, which can be canceled taking the following
ratio
\begin{equation}
\label{ratiophi}
\frac{g_{\phi \pi^+\pi^-}^{(s)}}{g_{\phi K^+ K^-}}
=\frac{g_{\Omega_8 \pi^+\pi^-}}{g_{\Omega_8 K^+ K^-}}.
\end{equation}   
Therefore, from eq. (\ref{fracgs}), one has
\begin{equation}
\label{gcouple}
g_{\phi \pi^+\pi^-}^{(s)}(s)=\frac{Q_{13}+Q_{23}}{Q_{11}+Q_{21}} 
\,g_{\phi K^+ K^-}(s),
\end{equation}
with $g_{\phi K^+ K^-}$ given in eq. (\ref{Lphikk}). Here we are
neglecting the mass difference between the $\Omega_8$ and the $\phi$ resonance,
which is around 100 MeV. In any case one has to take into account that:
 1) The important $\rho$ exchange effect is also canceled
in the ratios. 2) We have removed in eq. (\ref{Qij})
the three-momenta factors. As a result, the remaining differences coming from the mass
difference should be rather tiny.

Finally, by adding the above contribution with that of
eq.(\ref{g8phikk}), we find
\begin{equation}
\label{totg}
g_{\phi \pi^+ \pi^-}=
g_{\phi \pi^+ \pi^-}^{(\gamma)}+g^{(\phi \omega)}_{\phi \pi^+ \pi^-}
+g_{\phi \pi^+ \pi^-}^{(s)},
\label{gcontribs}
 \end{equation}    
which allows us to obtain the  $\phi\rightarrow\pi^+ \pi^-$
 decay width as
we did before only for the $g^{(\gamma)}_{\phi\pi^+ \pi^-}$ coupling.
In order to determine the sign of the interference in eq.(\ref{totg})
it is important to know the sign of $F_V\,G_V$ (see eqs.(\ref{Lphikk}) and (\ref{Lphipipi})).
We have taken $F_V\,G_V > 0$ since the $L_9$ chiral
parameter, whose main resonance contribution is given by
$F_V \, G_V/2M_\rho^2$ \cite{Ecker},
is positive and large. 

\subsection{The OZI rule violation}

The direct coupling $g_{\phi \pi^+\pi^-}$ violates the OZI rule.
This is clearly seen
in a quark picture when considering the $\phi$ as a pure $s\bar{s}$ state.
From the QCD Lagrangian one can see the OZI rule as
a prediction of the $1/N_c$ expansion, with $N_c$ the number of colors.
While the couplings of the decays which
do not violate the OZI rule are $O(1/N_c^{1/2})$ \cite{witten}, those
that violate the OZI rule are suppressed by an extra $1/N_c$.
In addition, meson loops are
suppressed by at least one power of $1/N_c$ \cite{witten}.
As a consequence, the
$g^{(s)}_{\phi\pi^+\pi^-}$ coupling given in eq. (\ref{gcouple}),
which is due to kaon loops, as discussed above, is
${\mathcal{O}}(1/N^{3/2})$.
Note, in contrast, that the $g_{\phi K^+ K^-}$ coupling, from eq.
(\ref{Lphikk}), is order $1/N_c^{1/2}$, since $f$ and $G_V$ are
$O(N_c^{1/2})$ and $M_\phi$ is order 1.

However, in quark model calculations \cite{isgur} the large
$N_c$ suppression of two intermediate meson states is considered
insufficient in order to
explain the experimental success of the OZI rule. The point is that in these
models the real parts of the two meson loop contributions to OZI violating
processes, although
large $N_c$ subleading, are found to be
much larger than they should be in order to explain the experimental success of
the OZI rule. The solution advocated by the authors is that a cancellation
among a very large number of intermediate states seems to operate.
This is illustrated via the example of $\omega-\rho$ mixing in \cite{isgur}.

Nevertheless, one should notice that
the real part of the two-meson loop is divergent,
and the remnant finite part depends upon the
regularization and renormalization schemes, apart,
of course, from the details of the dynamical model.
In refs.\cite{isgur,isgur2} this regularization is done including
several cut-offs within a quark flux tube model, having an
explicit scale dependence. In contrast, we have just included kaons and pions
as intermediate states and we have
renormalized such contributions making use of a cut-off
$\approx \Lambda_{\hbox{ChPT}}$. Still,
the physical quantities we calculate are scale independent and well defined,
since
any change in the cut-off would be reabsorbed by a change in the $L_i$
ChPT counterterms. Note
that, since we are making use of an effective field theory formalism,
the chiral Lagrangian counterterms should take into
account any other contribution from  more massive
intermediate states.
In our approach we use
ChPT up to ${\mathcal{O}}(p^4)$ and generate
higher orders through eq. (\ref{IAM}). In this way,
any other contribution coming from heavier
virtual intermediate states is reabsorbed in the final values of the
$L_i$ counterterms given in table 1.
At this point, our previous statement
about the fact that our result for the
$g^{(s)}_{\phi \pi^+\pi^-}$ coupling is due to kaon loops is meaningful only
because we have taken a natural value for the cut-off. For such value, the
contribution from graphs without kaon or pion loops,
which come just from the $L_i$ counterterms, is between one and two orders of
magnitude smaller than that of kaon loops. Comparing our work
with that of refs. \cite{isgur,isgur2}, we cannot tell exactly the size
of each separate contribution due to each state more massive than the kaons.
If each one of these contributions was large as it happens in refs. \cite{isgur,isgur2},
then, we would also be finding a cancellation.

In order to obtain further support for our arguments
about the kaon loop size, it is instructive to revisit, within
the IAM formalism, the kaon loop
contribution to $\omega\rightarrow\pi^+\pi^-$ that we estimated in sect.II.B.
Note that the value obtained for the $\omega-\rho$ mixing from
kaon loops in sect.II.B was dependent on
the regularization scale. In contrast, in the IAM
this dependence is canceled with that of the
chiral parameters $L_i$. In addition, the IAM respects unitarity
and accounts for isospin breaking not only in the loops (through different
masses of the charged and neutral kaons), but also in the
$\phi K^+K^-$ and $\phi K^0 \bar K^0$ couplings and the
$K \bar K\rightarrow \pi^+\pi^-$ amplitudes.

In order to reinterpret our results for the $\Omega_8\pi^+\pi^-$ coupling
in terms of an $\Omega_8-\rho$ mixing and compare with
sect.II.A., we write (see fig.7)
\begin{equation}
g_{\Omega_8\pi^+\pi^-}=\tilde \Theta_{\Omega_8\rho}g_{\rho\pi^+\pi^-}
\frac{1}{M^2_{\Omega_8}-M_\rho^2+i\,M_\rho \Gamma_\rho},
\end{equation}
with $g_{\rho\pi^+\pi^-}=-G_V s /(f^2 M_\rho^2)$ from eq.(\ref{Lag}).
This gives us $\tilde \Theta_{\Omega_8\rho}$, from where, using
eq.(\ref{mixing}) and the fact that the $\omega_1$ does
not couple to $K\bar K$ at the leading chiral order, we obtain
\begin{equation}
\tilde \Theta_{\omega\rho}=\frac{1}{\sqrt{3}}\tilde \Theta_{\Omega_8\rho}
=\frac{1}{\sqrt{3}}\frac{g_{\Omega_8\pi^+\pi^-}}{g_{\rho\pi^+\pi^-}}
\left[ M^2_{\Omega_8}-M_\rho^2+i\,M_\rho \Gamma_\rho \right].
\end{equation}
Taking now the value for $g_{\Omega_8\pi^+\pi^-}$ obtained
in the IAM from eq.(\ref{fing}), with
$g_{\Omega_8 K^+ K^-}=-\sqrt{3/2}g_{\phi K^+ K^-}$
from eq.(\ref{Lag}), we arrive at a value
of $\tilde\Theta_{\omega\rho}(M_\rho)=(-52-i76)\,\hbox{MeV}^2$
and $\tilde\Theta_{\omega\rho}(M_\phi)=(-299-i81)\,\hbox{MeV}^2$.
 These results corroborate the ``order of magnitude''
arguments given in sect.II.B., obtained using the non-unitary
eq.(\ref{loopkaon}), to show that the kaon loop contributions
are very small relative to the dominant OZI allowed contribution.

It is also interesting to remark that the cancellation between mesons
loops in the model of ref. \cite{isgur} does not operate for the scalar sector
with vacuum quantum numbers $J^{PC}=0^{++}$ as discussed in ref. \cite{isgur2}.
The failure of the large $N_c$ suppression
in this sector, and its associated OZI rule violation, is also discussed in more
general terms in ref. \cite{Stern}. Although the scalar sector is very hard to discuss in
terms of quark models, due to the large rescattering effects, it is equally well
described as the vector channels in the framework of non-perturbative unitarity
methods from the ChPT series \cite{oop,ollerpaco,oo,npa,gamma}, see 
also fig. 5. For instance,
in refs. \cite{npa,gamma} the $\sigma$, $f_0(980)$ and $a_0(980)$ were dynamically generated and
their meson--meson and $\gamma \gamma$ decay modes were 
analyzed in very good agreement with
experiment. Furthermore, in ref. \cite{oo} the spectrum in the scalar sector was discussed
taking into account as well the large $N_c$ limit. In addition, the presence of a scalar nonet
due to the meson-meson self interactions, which disappears in the limit
$N_c \rightarrow \infty$, was then established. On the other hand, it was
also found that the
lightest preexisting scalar nonet, with mass ${\mathcal{O}}(1)$ in the $N_c$ counting,
should comprise a singlet around 1 GeV and an octet around 1.4 GeV, in qualitative agreement
with the expectations of ref. \cite{isgur2}. The success of our approach in the
$0^{++}$ sector indicates that our techniques are powerful in the study of OZI
violating processes. Note that we describe both vector and scalar channels
  without including any new ad-hoc elements.
 
\section{ Results and discussion }

\subsection{$\phi\rightarrow\pi^+\pi^-$ Branching Ratio}

In this section we 
are going to present the 
resulting branching ratios for the $\phi \rightarrow \pi^+\pi^-$ decay. 
To do that we will consider and discuss the different sources 
contributing to the total 
$g_{\phi\pi^+\pi^-}$ coupling as given in eq.(\ref{gcontribs}).

We first consider the contribution 
$g_{\phi \pi^+\pi^-}^{(\gamma)}$ introduced in 
sec.II.A. We take as a final value $g_{\phi\pi^+\pi^-}^{(\gamma)}
\simeq [10.6\pm0.4-i\,(4.47\pm0.15)]10^{-3}$ where the 
uncertainty
is mainly due to the value 
of $F_V$, which ranges between $F_V=154$ MeV, coming
from the $\rho\rightarrow e^+e^-$ decay, and
$F_V=165$ MeV, coming
 from the $\phi \rightarrow e^+e^-$ decay, when evaluating
both them with eqs. (\ref{Lag}) and (\ref{Vtilde}).

Concerning the kaon-loop contributions to the $\phi-\rho$ 
mixing eq.(\ref{gcouple}),
after averaging over all the fits presented in table 1, we obtain:
\begin{eqnarray*}
g_{\phi\pi^+\pi^-}^{(s)} \simeq -[5.6\pm 0.4-i\,(3.8\pm0.12)] 10^{-3}.
\end{eqnarray*}
Let us note that the error is mainly due to
 the differences between the $L_i$ corresponding to the different
 fits, since they are  much
larger than the errors given by MINUIT, which are certainly 
underestimated. Furthermore,
we have checked that this error band spans the dispersion in the results
due to the variations of the chiral parameters that could 
yield a reasonable fit.

Although they were not present in eq.(\ref{gcontribs})
there are corrections coming from
diagrams with photon loops which are expected to be of
the same order of magnitude as the isospin breaking corrections
from the different mass of charged and neutral mesons
\cite{knecht,ulf1,ulf2,gasser}.
 We do not have means at present to evaluate these diagrams within the
  non-perturbative  chiral scheme which we have followed. One would
  also need counterterms whose values are nowadays unknown.
However, explicit calculations of the absorptive part of the
$\eta\gamma$ intermediate channel in ref. \cite{bramon}
give a contribution of, {\it at most},
1/4 of the kaon loops but {\it with opposite sign}. 
This $\eta\gamma$ will be our largest source 
of uncertainty in the errors given for 
each one of the different $\phi-\omega$ scenarios, that we 
discuss next.

As we have already commented, the contribution from the two step
$\phi-\omega-\rho$ mechanism, depends on the $\phi-\omega$ mixing.
Our results are the following:

\begin{itemize}
\item Strong scenario: we find 
$g_{\phi\pi^+\pi^-}^{\omega\phi}=[4.4-i3.7]10^{-3}$ or
  $g_{\phi\pi^+\pi^-}^{\omega\phi}=[6.0-i5.6]10^{-3}$,
  depending on whether we use
  $\hbox{Re}\,\tilde\Theta_{\phi\omega}=20000$ or $29000\, \hbox{MeV}^2$,
  respectively. Therefore, there is a large cancellation with the kaon 
  loop contribution, and we obtain:
\begin{eqnarray*}
BR\simeq (1.7\pm0.3)\times10^{-4}\;\hbox{to} \; (2.5\pm0.3)\times10^{-4},
\end{eqnarray*}
where the uncertainty in the central values depends on whether we use
$\hbox{Re}\,\tilde\Theta_{\phi\omega}=20000$ or $29000\, \hbox{MeV}^2$, 
respectively.

\item Weak  scenario: we get $g_{\phi\pi^+\pi^-}^{\omega\phi}=[-0.73-i
  0.61]10^{-3}$, very small compared with both the electromagnetic and
  kaon-loop contributions. Thus, there is only a partial cancellation of the 
electromagnetic contribution with that of kaon loops, and we obtain
\begin{eqnarray*}
BR\simeq (0.38\pm0.12)\times10^{-4}.
\end{eqnarray*}

\end{itemize}

Apart from the contributions discussed so far, 
there is also the possibility of local 
terms giving rise to
a direct $\rho-\phi$ mixing.
 However, one can argue that, 
by resonance saturation, 
the inclusion of the two-step process $\phi-\omega-\rho$ can 
be enough to take care 
of such local terms by considering that they are
resummed on the $\omega$ propagator.


\section{Conclusions}

In this work we have evaluated the kaon loop contribution to the
$\phi\rightarrow \pi^+\pi^-$ decay via $\phi-\rho$ mixing
from the splitting of meson
  masses, making use of the
unitarized chiral amplitudes with strong
  isospin breaking.
We have shown that although
 this strong contribution to the $\phi \rightarrow
  \pi^+\pi^-$ decay  gives rise to smaller
branching ratios by itself than the tree level
electromagnetic contributions, they can have a very large destructive
interference with either the electromagnetic or the 
$\phi-\omega-\rho$
contributions.

  We have also estimated the error in our $\phi \rightarrow \pi^+\pi^-$ 
branching ratio
  calculation coming from the uncertainties in
  $F_V$, the fitted ${\mathcal{O}}(p^4)$ 
ChPT counterterms, the photon-loop contributions,
as well as the considered
$\phi-\omega$ mixing scenarios.

A complete calculation of the loops with photons
is  missing in the present
work, although they have been estimated making use of the
results of ref. \cite{bramon}. Still, they are the main 
source of uncertainty within each $\phi-\omega$ mixing scenario.

Accepting this additional uncertainty,
we find that the strong coupling scenario \cite{achasov} yields
\begin{eqnarray*}
BR\simeq (1.7\pm0.3)\times10^{-4}\;\hbox{to} \;
(2.5\pm0.3)\times10^{-4},
\end{eqnarray*}
in very good agreement with the experimental results of ref.\cite{CMD2}.
In contrast, the Weak \cite{achasov} 
scenario yields
\begin{eqnarray}
BR\simeq (0.38\pm0.12)\times10^{-4},\nonumber
\end{eqnarray}
It seems to
prefer a value somewhat lower
 than the experimental value provided by ref.\cite{SND}, 
although still reasonably compatible with it.

Of course, a precise determination of the photon
loops in the non-perturbative regime would be desirable
to reduce the theoretical uncertainties.

Finally, we would like to remark that the solution of the
experimental conflict in the $\phi\rightarrow\pi^+\pi^-$ will,
eventually, help us
to discard some of the $\phi-\omega$ mixing scenarios
proposed in the literature.

\section*{Acknowledgments}
We would like to thank T. Barnes, A. Bramon, R. 
Escribano and Ulf-G. Mei{\ss}ner
 for discussions and useful information. J.A.O.
would like to acknowledge partial financial  support from
  the Generalitat Valenciana. J.R.P. thanks the Departamento de F\'{\i}sica
Te\'orica and IFIC at the University of Valencia--CSIC for their 
warm hospitality.
We would also like to acknowledge
  financial support from the DGICYT under contracts PB96-0753,
  AEN97-1693 and PB98-0782 and from the EU TMR network Eurodaphne, contract no.
  ERBFMRX-CT98-0169.

\appendix
\section{Amplitudes}

In this appendix we give the expression for the $J=1$ partial waves
obtained from the ChPT Lagrangian, but setting $m_u\neq m_d$. The normalization
of the $T$-matrix used here is the same as in ref. \cite{oop}. Let us first
define the modulus of the CM momenta of the different particles as
\begin{eqnarray*}
p_{\pi^+}=\sqrt{\frac{s}{4}-m_{\pi^+}^2},
p_{K^+}=\sqrt{\frac{s}{4}-m_{K^+}^2},
p_{K^0}=\sqrt{\frac{s}{4}-m_{K^0}^2}.
\end{eqnarray*}
where $m_{\pi^+}$ is the charged pion mass. Then, once they are projected in
$P$-wave, the tree level amplitudes from the $O(p^2)$and $O(p^4)$ Lagrangian for
$K^+K^-\rightarrow\pi^+\pi^-$ scattering are

\begin{eqnarray*}
T_2(s,t,u)&=&-\frac{p_{\pi^+} p_{K^+}} {3\,f_{K^+}\,f_{\pi}} , \\
T^P_4(s,t,u)&=&\frac{4}{3\,f_{K^+}^2\,f_{\pi}^2}
\left[L_3\,s -L_5 (m_{K^+}^2+m_{\pi^+}^2)\right]\, p_{\pi^+} p_{K^+},
\end{eqnarray*}
whereas for $K^0\bar{K}^0\rightarrow\pi^+\pi^-$ scattering they are given by

\begin{eqnarray*}
T_2(s,t,u)&=&\frac{p_{\pi^+} p_{K^0}} {3\,f_{K^0}\,f_\pi}, \\
T^P_4(s,t,u)&=&-\frac{4}{3\,f_{K^0}^2\,f_{\pi}^2}
\left[ L_3\,s- L_5 (m_{K^0}^2+m_{\pi^+}^2)\right]\, p_{\pi^+} p_{K^0}.
\end{eqnarray*}
In the above formulas,
$f_\pi$, $f_{K^+}=1.22\,f_\pi$ and $f_{K^0}$ are the decay constants of the
charged pion, kaon and neutral kaon, respectively. In the approach we are
following here of neglecting tadpoles one has, up to ${\mathcal{O}}(p^4)$, that

\begin{eqnarray*}
f_{K^0}=f_{K^+}(1+4 L_5\frac{m_{K^0}^2-m_{K^+}^2}{f_\pi^2} ).
\end{eqnarray*}
For $K^+K^-\rightarrow K^+K^-$ we obtain

\begin{eqnarray*}
T_2(s,t,u)&=&-\frac{2}{3\,f_{K^+}^2} \, p_{K^+}^2,  \\
T^P_4(s,t,u)&=&\frac{4 \, p_{K^+}^2}{3\,f_{K^+}^4}
\left[ 2(2L_1-L_2+L_3)s-4(2L_4+L_5)m_{K^+}^2\right],
\end{eqnarray*}
the $ K^0\bar{K}^0\rightarrow K^0\bar{K}^0$ amplitude is exactly the same,
but changing $m_{K^+}$ by $m_{K^0}$ and $f_{K^+}$ by $f_{K^0}$.
For $\pi^+\pi^-\rightarrow \pi^+\pi^-$, we find
\begin{eqnarray*}
T_2(s,t,u)&=&-\frac{2} {3\,f_\pi^2}\, p_{\pi^+}^2, \\
T^P_4(s,t,u)&=&\frac{8\, p_{\pi^+}^2}{3\,f_{\pi}^4}
\left[(2L_1-L_2+L_3)\,s-(4L_4+2L_5) m_{\pi^+}^2\right].
\end{eqnarray*}
We have left the $K^+K^-\rightarrow K^0\bar{K}^0$ amplitude for
the end, since we had an erratum in our previous paper \cite{oop}.
Thus, we first give the {\em complete amplitude in the
isospin limit}, before projecting on the
P-wave. It reads
\begin{eqnarray*}
T_2(s,t,u)&=&\frac{u-2m_K^2}{2f_K^2},\\
T_4(s,t,u)&=&\frac{-2}{f_K^4}\left[
(4L_1+L_3)(s-2m_K^2)^2+
2L_2(u-2m_K^2)^2\right.\\
&&+\,(2L_2+L_3)(t-2m_K^2)^2+
8m_K^4(L_8+2L_6)\\
&&-\,\left.2um_K^2L_5-8m_K^2(2m_K^2-s)L_4\right].
\end{eqnarray*}
The P-wave in the isospin breaking case is given by:
\begin{eqnarray*}
\label{kpkmK0K0}
T_2(s,t,u)&=&-\frac{p_{K^+} p_{K^0}}{3\,f_{K^+}\,f_{K^0}} , \\
T^P_4(s,t,u)&=&\frac{4\, p_{K^+} p_{K^0}}{3\,f_{K^+}^2\,f_{K^0}^2}
\left[L_3\,s- L_5 (m_{K^+}^2+m_{K^0}^2) \right].
\end{eqnarray*}

Finally, we give the loop function $G=diag(g_1,g_2,g_3)$,
where $g_i$ is 
\begin{eqnarray*}
g_i(s)=\frac{1}{(4\pi)^2}\left[\sigma_i\log\frac{\sigma_i\,Q_i+1}
{\sigma_i\,Q_i-1}-2\log\left(\frac{q_{max}}{m_i}(1+Q_i)
\right)\right].
\end{eqnarray*}
where $\sigma_i(s)=\sqrt{1-4m^2_i/s}$ and $Q_i=\sqrt{1+m_i^2/q_{max}^2}$.

\onecolumn

\begin{figure}[htb]
\begin{center}
\begin{picture}(25000,7500)
\THICKLINES
\bigphotons
\drawline\fermion[\E\REG](4500,4500)[5000]
\drawline\fermion[\E\REG](4500,4000)[5000]
\drawline\photon[\E\REG](\pbackx,\pbacky)[4]
\put(9500,4250){\circle*{1000}}
\global\advance \pbackx by 100
\global\advance \pbacky by -50
\drawline\fermion[\SE\REG](\pbackx,\pbacky)[4000]
\global\advance \pfronty by  200
\drawline\fermion[\NE\REG](\pfrontx,\pfronty)[4000]
\put (6500,5500) {\Large $\phi$}
\put (11500,5500){\Large $\gamma$}
\put (17000,7000) {\Large $\pi^+$}
\put (17000,700) {\Large $\pi^-$}
\end{picture}          
\end{center}      
\caption{$\phi\rightarrow\pi^+\pi^-$ decay through a photon.}            
\end{figure}
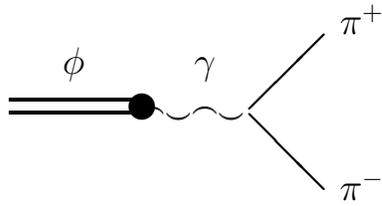

\begin{figure}[htbp]
\begin{center}
\hspace*{-1.7cm}
\hbox{\psfig{file=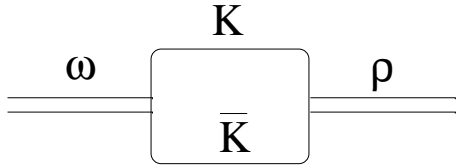,width=6cm,angle=-90}}    
\end{center}
\caption{Kaon loop contribution to the $\rho-\omega$ mixing.}
\end{figure}

\begin{figure}[ht]
\centerline{
\protect
\hbox{
\psfig{file=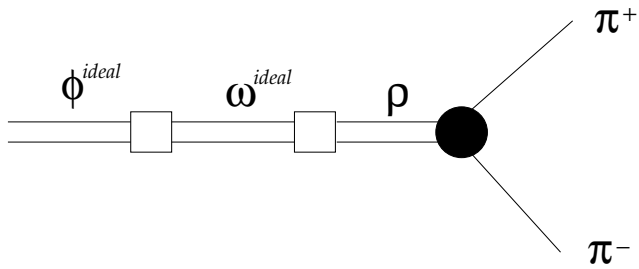,width=0.6\textwidth}}}
\caption{Two step mechanism for $\phi\rightarrow\pi^+\pi^-$ decay.}
\end{figure}

\begin{figure}[htbp]
\begin{center}
\begin{picture}(40000,8000)

\THICKLINES
\bigphotons

\drawline\fermion[\E\REG](1500,5000)[6000]
\drawline\fermion[\E\REG](1500,4500)[6000]
\put(9500,4750){\circle{6000}}
\global\advance \pbackx by 4000
\drawline\fermion[\SE\REG](\pbackx,\pbacky)[4000]
\drawline\fermion[\NE\REG](\pfrontx,\pfronty)[4000]
\put (4000,6000) {\Large $\phi$}
\put (9000,7500) {\large $K^+$}
\put (9000,1000) {\large $K^-$}
\put (15000,8000) {\Large $\pi^+$}
\put (15000,1000) {\Large $\pi^-$}

\drawline\fermion[\E\REG](22500,5000)[6000]
\drawline\fermion[\E\REG](22500,4500)[6000]
\put(30500,4750){\circle{6000}}
\global\advance \pbackx by 4000
\drawline\fermion[\SE\REG](\pbackx,\pbacky)[4000]
\drawline\fermion[\NE\REG](\pfrontx,\pfronty)[4000]
\put (25000,6000) {\Large $\phi$}
\put (29500,7500) {\large $K^0$}
\put (29500,1000) {\large $\bar{K}^0$}
\put (36300,7500) {\Large $\pi^+$}
\put (36300,1000) {\Large $\pi^-$}
\end{picture}          
\caption{Kaon loop contributions to the 
$\phi\rightarrow \pi^+\pi^-$ decay. If the
charged and neutral kaons had the same mass,
the two diagrams would cancel.}
   \end{center}
\end{figure}
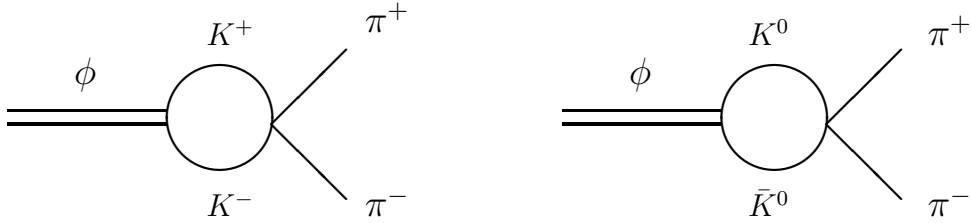

\begin{figure}[htbp]
\begin{center}
\hspace*{-1.7cm}
\hbox{\psfig{file=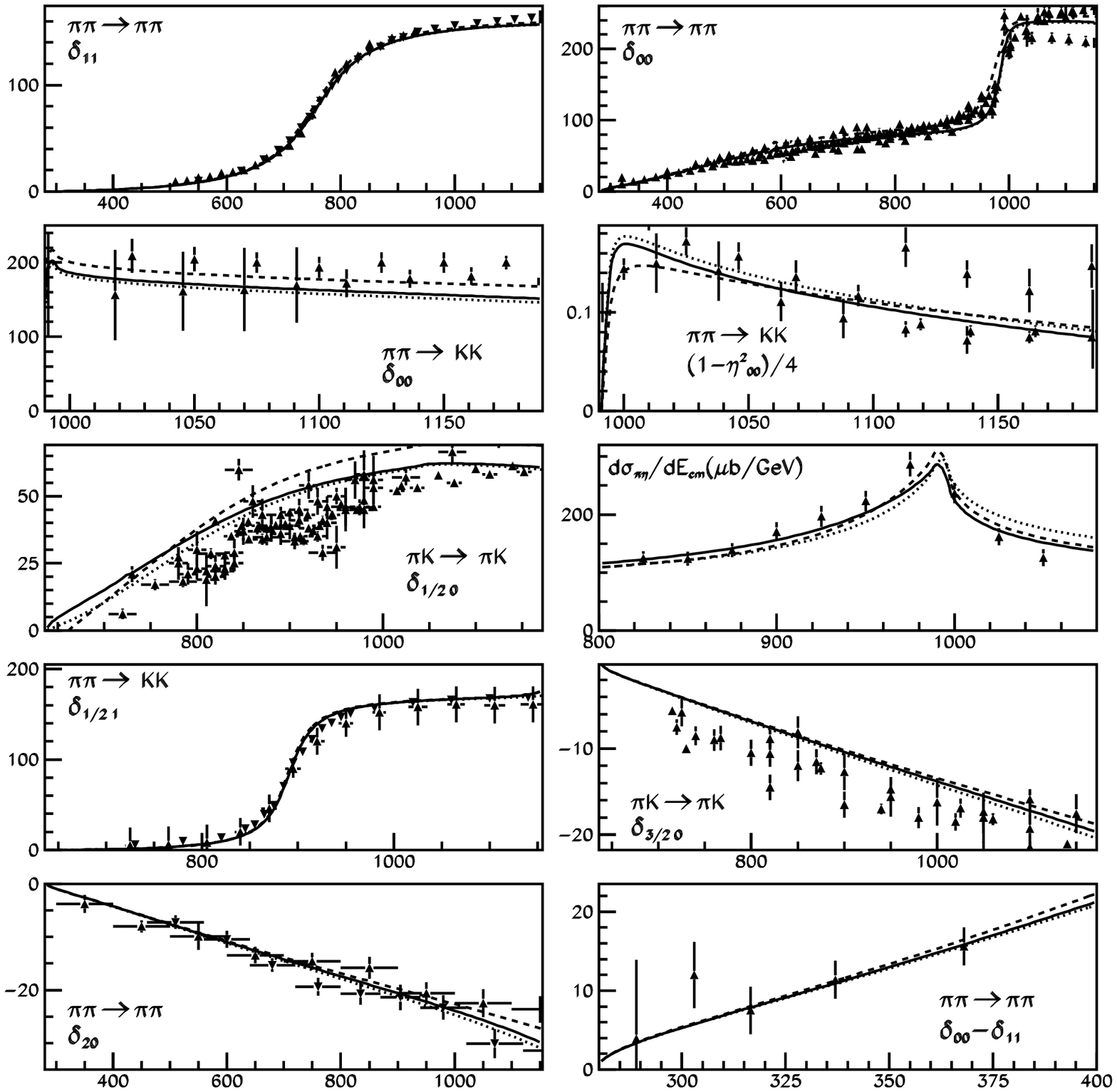,width=18cm}}    
\end{center}
\caption{\label{fig:result}}
{\small Coupled channel IAM results for meson-meson scattering.
The dashed, continuous and dotted lines are obtained, respectively,
with the chiral parameter sets 1, 2 and 3 given in table 1.
Note that they are indistinguishable for almost every channel.
The experimental data for each plot, starting from left to right
and top to bottom, comes from \cite{Prot,Esta},
\cite{Prot,d00,Frog,exp}, \cite{Ma,eta},
\cite{Frog,Ma,eta}, \cite{Mercer,kk012,Esta2},
\cite{Amsterdam}, \cite{Mercer,Esta2}, \cite{Esta2,Linglin},
\cite{schenk,roselet2} and, finally \cite{roselet}}
\end{figure}

\begin{table}[htbp]
{ \setlength{\tabcolsep}{.3mm}
\begin{tabular}{|c|c|c|c|c|c|c|c|c|}
\hline
\small{
Fit} & \small{$\hat L_1$} & \small{$\hat L_2$} & \small{$\hat L_3$} &
\small{$\hat L_4$} & \small{$\hat L_5$} &
\small{$2\hat L_6+\hat  L_8$} & \small{$\hat L_7$} &
\begin{tabular}{c}
 \small{$q_{max}$}\\
\small{(MeV)}
\end{tabular}
\\  \hline
\small{set 1}&0.91&1.61&-3.65&-0.25&1.07&0.58&-0.4&666 
\\ \hline
\small{set 2}&0.91&1.61&-3.65&-0.25&1.07&0.58&0.05&751 
\\ \hline
\small{set 3}&0.88&1.54&-3.66&-0.27&1.09&0.68&0.10&673 
\\ \hline\hline
& \small{$L_1$} & \small{$L_2$} & \small{$L_3$} & \small{$L_4$} & \small{$L_5$}&
 \small{$2L_6+L_8$} & \small{$L_7$} & \small{$\mu$}
 \\ \hline
\small{ChPT}&0.4&1.4&-3.5&-0.3&
1.4&0.5&-0.4&$M_\rho$
\\
\cite{cual}&$\pm$0.3&$\pm$0.3&$\pm$1.1&$\pm$0.5&$\pm$0.5&$\pm$0.3
&$\pm$0.2&\\ \hline
\end{tabular}
}
\caption{\label{tabla2}}
{\small Different sets of chiral parameters (in $10^{-3}$ units)
that yield reasonable fits to
the meson-meson scattering phase shifts. We have
used a hat to
differentiate them from those obtained within standard ChPT 
\cite{cual}, since in
our case we have already differences at the ${\mathcal{O}}(p^4)$
 with respect the 
next-to-leading ChPT amplitudes and we have used high
energy data in the fit. However, as it is explained in the text, we still
expect them to be relatively similar once the scales are chosen appropriately
(roughly $\mu\simeq1.2\,q_{max}$, see \cite{oop} for details).}
\end{table}

    \begin{figure}[htbp]
      \begin{center}
\begin{picture}(35000,8000)

\THICKLINES
\bigphotons
\put (7000,500) {(a)}
\drawline\fermion[\SE\REG](1600,7500)[4000]
\drawline\fermion[\NE\REG](1600,2000)[4000]

\drawline\fermion[\E\REG](4500,5000)[6000]
\drawline\fermion[\E\REG](4500,4500)[6000]

\global\advance \pbacky by 200

\drawline\fermion[\SE\REG](\pbackx,\pbacky)[4000]
\drawline\fermion[\NE\REG](\pfrontx,\pfronty)[4000]

\put (7000,6000) {\Large $\Omega_8$}

\put (13500,8000) {\large $K^+$}
\put (13500,1000) {\large $K^-$}
\put (100,8000) {\large $K^+$}
\put (100,1000) {\large $K^-$}

\put (27000,500) {(b)}

\drawline\fermion[\SE\REG](21600,7500)[4000]
\drawline\fermion[\NE\REG](21600,2000)[4000]

\drawline\fermion[\E\REG](24500,5000)[6000]
\drawline\fermion[\E\REG](24500,4500)[6000]

\global\advance \pbacky by 200

\drawline\fermion[\SE\REG](\pbackx,\pbacky)[4000]
\drawline\fermion[\NE\REG](\pfrontx,\pfronty)[4000]

\put (27000,6000) {\Large }

\put (27000,6000) {\Large $\Omega_8$}

\put (32500,8000) {\large $\pi^+$}
\put (32500,1000) {\large $\pi^-$}
\put (21000,8000) {\large $K^+$}
\put (21000,1000) {\large $K^-$}
\end{picture}         
        \caption{$K^+K^- \rightarrow K^+ K^-$
and  $K^+ K^- \rightarrow\pi^+ \pi^-$ processes occurring through the
exchange of an $\Omega_8$.}
        \label{fig:diagramas}
      \end{center}
\end{figure}
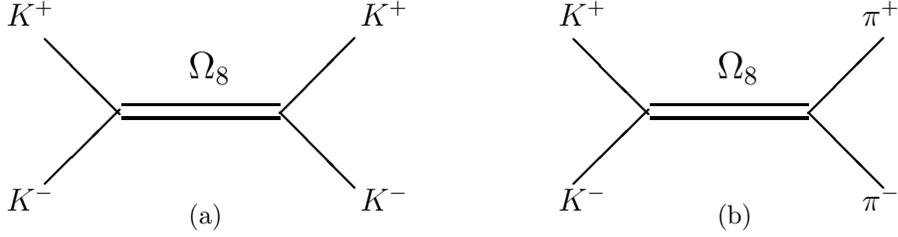

\begin{figure}[htbp]
\begin{center}
\hspace*{-1.7cm}
\hbox{\psfig{file=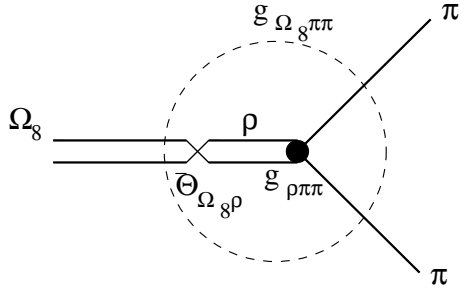,width=6cm,angle=0}}    
\end{center}
\caption{
The $g_{\Omega_8\pi^+\pi^-}$ coupling interpreted
as a $\Omega_8-\rho$ mixing and a $\rho\rightarrow\pi^+\pi^-$ decay.}
\end{figure}

\end{document}